\documentclass[journal]{IEEEtran}
\usepackage{ifpdf}

\usepackage{cite}

\ifCLASSINFOpdf
   \usepackage[pdftex]{graphicx}
  \graphicspath{{figs/}{../pdf/}}
\else
\fi

\usepackage{multirow}
\begin{document}
%
\title{Studies on the Cherenkov Effect for Improved Time Resolution of TOF-PET}
%
%
%

\author{{S. E. Brunner, L. Gruber, J. Marton, K. Suzuki, and A. Hirtl}%
\thanks{Manuscript received May 24, 2013. This work was supported in part by EU-project HadronPhysics3 (project 283286).}%
\thanks{S.~E.~Brunner (email: stefan.enrico.brunner@oeaw.ac.at), L.~Gruber, J.~Marton and K.~Suzuki are with the Stefan-Meyer-Institute for Subatomic Physics of the Austrian Academy of Sciences, Vienna, Austria. A.~Hirtl is with the Department of Biomedical Imaging and Image-guided Therapy of the Medical University of Vienna, Austria.}}%
\maketitle
\begin{abstract}
With the newly gained interest in the time of flight method for positron emission tomography (TOF-PET), many options for pushing the time resolution to its borders have been investigated. As one of these options the exploitation of the Cherenkov effect has been proposed, since it allows to bypass the scintillation process and therefore provides almost instantaneous response to incident 511\,keV annihilation photons. Our simulation studies on the yield of Cherenkov photons, their arrival rate at the photon detector and their angular distribution reveal a significant influence by Cherenkov photons on the rise time of inorganic scintillators - a key-parameter for TOF in PET. A measurement shows the feasibility to detect Cherenkov photons in this low energy range.
\end{abstract}


%
\IEEEpeerreviewmaketitle

%
%
%
%

 




\section{Introduction}
\IEEEPARstart{I}{n} recent years, the Cherenkov effect for electrons at energies below 511\,keV has become subject of investigations for improving the time resolution of time of flight positron emission tomography (TOF-PET) \cite{Lecoq10,Dolenec10}. The extent of improvement in coincidence time resolution (CTR) of PET and, thus, in signal-to-noise ratio (SNR) is promising and has been investigated in detail in ref. \cite{Brunner13}.

In inorganic scintillators, optical photons are emitted following the interaction of a 511\,keV annihilation photon with the scintillator, leaving an inner shell hole and an energetic primary electron, followed by a cascade of energy relaxation processes: radiative (secondary X-rays) and non-radiative decay (Auger processes), inelastic electron-scattering in the lattice, thermalization, electron$-$phonon interactions, trapping of electrons and holes and energy transfer to luminescent centers. All of them are introducing additional time spread to the emission of scintillation photons \cite{Derenzo03, Williams01}. Most of these processes are irrelevant for the Cherenkov photons, since their emission takes place in the early stages of the relaxation cascade (in the phase of electron scattering) and, thus, provide a more precise time stamp compared to scintillation photons.

The kinetic energy of electrons after photoelectric interaction with 511\,keV photons is dependent on their binding energy in the material and ranges from about 450\,keV\,-\,510\,keV, being sufficiently high for the emission of Cherenkov photons.

A short rise time is one key-parameter for good time resolution of scintillators and is subject of investigations in TOF-PET \cite{Lecoq10,Moses99,Derenzo03,Shao07}. This work focuses on the time distribution of both, scintillation and Cherenkov photons, i.e., their creation time inside the crystal and their arrival time at the photon detector.

In the following, results of calculations and Geant4 \cite{Geant4} simulations on the yield of Cherenkov photons, their angular distribution, their influence on the observable rise time and their absorption inside the scintillators will be shown. Finally, results of measurements of a basic coincidence setup using lead glass as emitter of Cherenkov radiation will be presented.

\section{Simulation}
\label{sec:simulation}

\begin{figure}[hbt]
 \centering
\includegraphics[width=0.7\columnwidth,keepaspectratio]{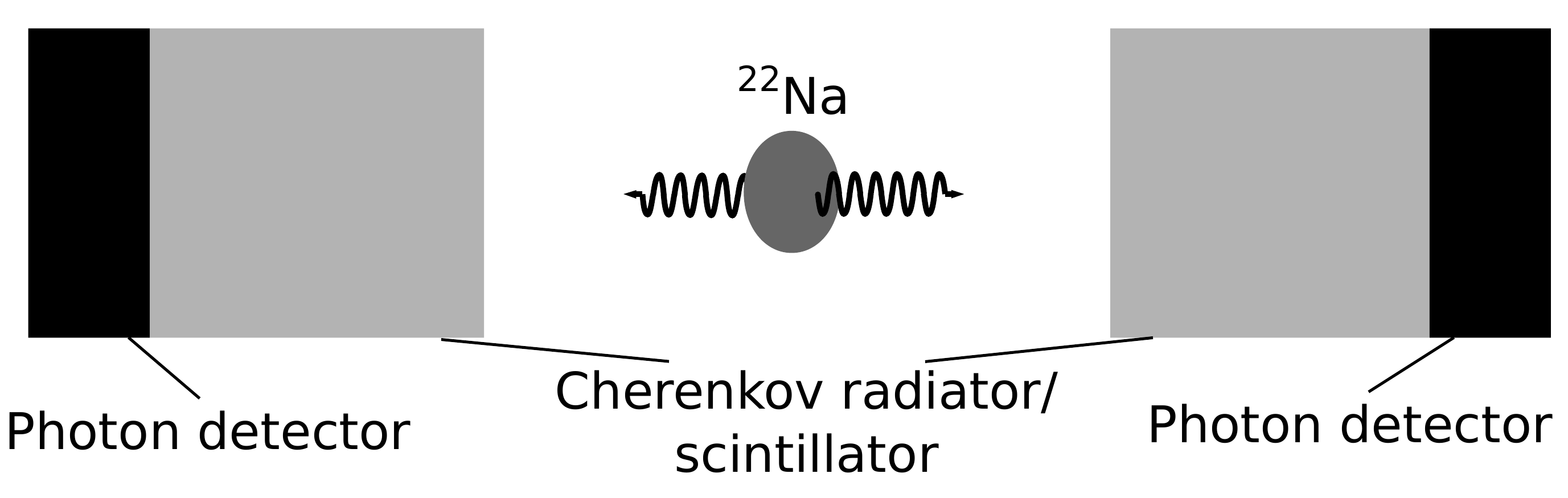}
\caption{Basic coincidence setup used for the Geant4 simulation studies. The Cherenkov radiators/scintillators have a size of 3\,mm\,$\times$\,3\,mm\,$\times$\,3\,mm, the photon detector attached has a sensitive surface of 3\,mm\,$\times$\,3\,mm.}
\label{fig:setup}
\end{figure}

\begin{table}[!t]
	\centering
	\caption{Input values for calculations and simulations.}
	\label{tab:inputvalues}
	\begin{tabular}{lcccc}
	\hline \hline
	  Material&Density [g/cm$^3$]&n&$\lambda_1$ [nm]&LY [photons/MeV]\\ \hline
	  LSO:Ce&7.4&1.82&390&27300\\
	  LuAG:Ce&6.7&1.84&260\footnotemark[1]&14000\\
	  BGO&7.13&2.15&310&8000\\
	  PWO&8.28&2.2&340&210\\
	  Pb-glass&5.05&1.79&340\footnotemark[2]&-\\ \hline \hline
	\end{tabular}
\end{table}
\footnotetext[1]{For transmission the wavelength bands ($\lambda_1$\,-\,$\lambda_2$) 260\,nm\,-\,320\,nm, 360\,nm\,-\,420\,nm and 480\,nm\,-\,1000\,nm are used ($\lambda_1$ is the lower, $\lambda_2$ is the upper cutoff wavelength).}
\footnotetext[2]{Estimated value.}

The simulations were performed with Geant4, v9.4. p3, using the Geant4-Livermore libraries for electromagnetic processes. The simulated crystals are cerium doped lutetium-oxyorthosilicate (LSO:Ce), cerium doped lutetium-aluminum-garnet (LuAG:Ce), lead-tungstate (PWO), bismuth-germanate (BGO) and lead glass with a cubic shape and edge lengths of 3\,mm. Their surfaces were polished and surrounded by air. For optical photon detection, a photon detector with a size of 3\,mm\,$\times$\,3\,mm was attached to one of the faces of the crystals. With these geometries, simple coincidence setups were simulated with $^{22}$Na as source of 511\,keV photons, see figure \ref{fig:setup}. The photon detectors were assumed to be ideal, i.\,e., infinite time resolution and a photon detection efficiency of 1. The creation time of the annihilation photons with 511\,keV represents time $t=0$ for the simulation. Due to ambiguous numbers in the literature originating from various experiments \cite{Derenzo00, Seifert12}, 100 ps were 
assumed for the rise time of all scintillators. The input values for the refractive index, $n$, the lower cutoff wavelength of the transmission spectrum, $\lambda_1$, and the light yield (LY), are given in table \ref{tab:inputvalues} \cite{Mao07, Auffray09, Dafinei95, Salacka10, Melcher92}.

In the following, the scintillation yield is the number of optical photons emitted by scintillation, the Cherenkov yield is the number of optical photons emitted due to the Cherenkov effect and an event is the interaction of a 511\,keV photon in the crystal by the photoelectric effect.

\subsection{Yield of Cherenkov Photons}
\label{sec:cherenkovyield}
The number of Cherenkov photons, N, emitted by an electron traveling faster than the speed of light in a dielectric medium can be calculated using,
\begin{equation}
\label{eq:NbCherPhot}
 \frac{dN^2}{dxd\lambda}=\frac{2\pi \alpha}{\lambda^2}\cdot \left(1-\frac{1}{\beta^{2}n^{2}(\lambda)}\right),
\label{eq:cheryield}
 \end{equation}
with x, being the electron range, $\alpha$, being the fine structure constant, $\beta$, the electron velocity over the speed of light $v/c$ and $n$, the refractive index, which was assumed to be constant for all wavelengths \cite{Leo94}. The electron ranges in the materials were calculated by estimating an energy window in which Cherenkov photons are emitted during the propagation of the electrons. The upper energy threshold was estimated by subtracting the binding energy of an electron in the K-shell of the heaviest element of the material from the initial photon energy of 511\,keV (binding energies were taken from ref. \cite{Thompson09}). Electrons in the K-shell have the maximum cross-section for interaction with 511\,keV photons by the photoelectric effect. The lower limit of the energy window, $E_\mathrm{thr}$ was determined by,
\begin{equation}
 E_\mathrm{thr} = m_\mathrm{e}c^2\cdot\left(\frac{1}{\sqrt{1-\beta_\mathrm{t}^2}}-1\right)
\end{equation}
with the Cherenkov threshold $\beta_\mathrm{t}=1/n$. For the upper and the lower threshold, the electron ranges were calculated using the values from the NIST-Estar database \cite{NIST-Estar}. Subtracting the lower range from the upper range and using eq. \ref{eq:NbCherPhot} gives an estimate for the Cherenkov yield.

The numbers in table \ref{tab:nbcherphotons} give the expected yield of Cherenkov photons after the interaction of 511\,keV photons with the material by the photoelectric effect. Due to the fact that the scintillation yield of PWO and Pb-glass is low and zero, respectively, a separation of Compton scattered photons and photons which interacted by the photoelectric effect was not possible. Therefore, the numbers for these materials include both, the number of Cherenkov photons after Compton and photoelectric interaction. This results in an underestimation of the simulated number of Cherenkov photons.

\begin{figure}[bt]
 \centering
\includegraphics[width=1\columnwidth,keepaspectratio]{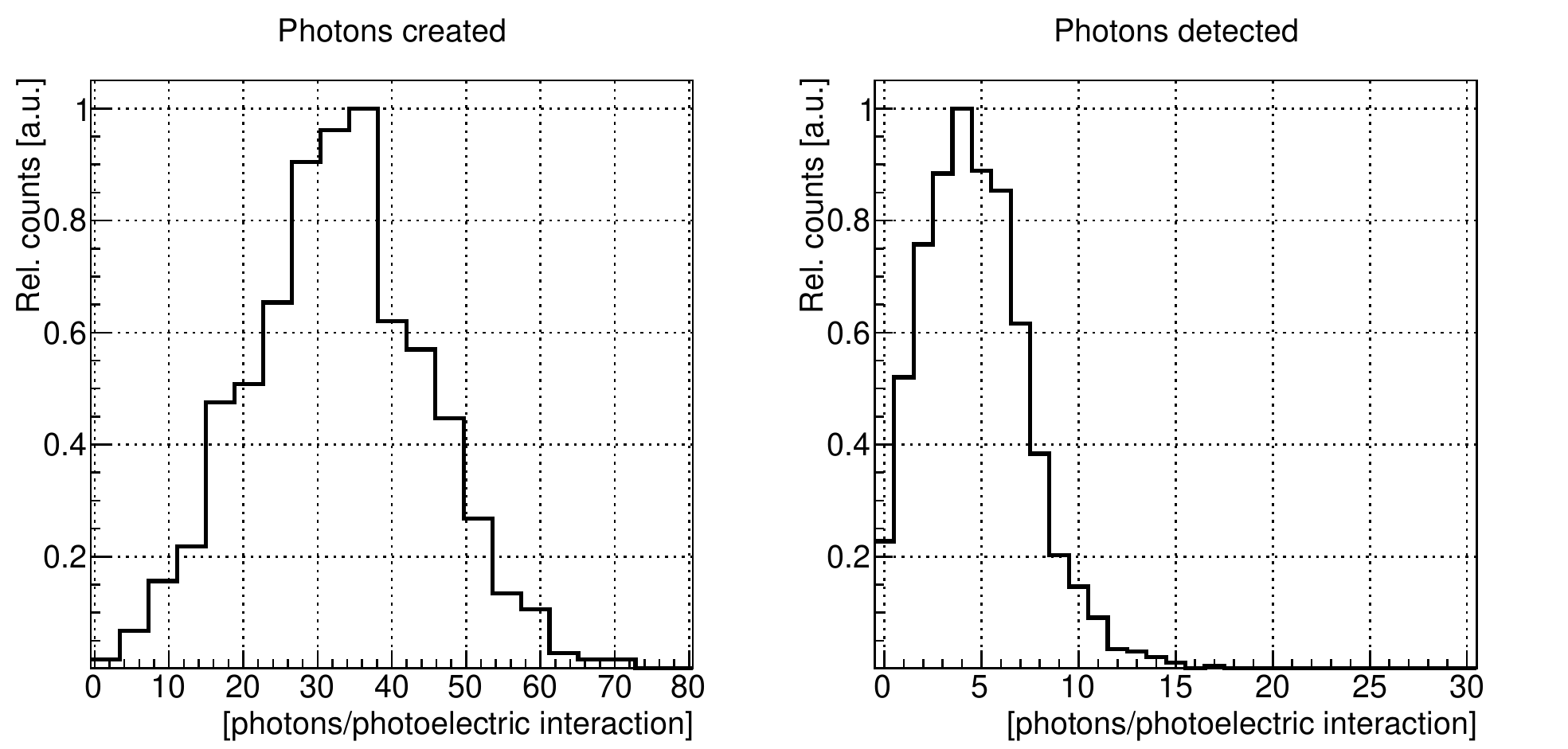}
\caption{Left: number of Cherenkov photons created after photoelectric interaction of 511\,keV photons with a cube of BGO with 3\,mm edge length. Right: number of detected Cherenkov photons with a photon detector of 3\,mm\,$\times$\,3\,mm, attached to the cube.}
\label{fig:nbcherphotons}
\end{figure}

An example of the distribution of created Cherenkov photons in a cube of BGO, emitted after photoelectric interaction of 511\,keV annihilation photons in the crystal, can be seen in figure \ref{fig:nbcherphotons}, on the left-hand side. On the right-hand side of figure \ref{fig:nbcherphotons}, the number of detected Cherenkov photons arriving at the photon detector, attached to the cube, can be seen.

\begin{table}[!t]
	\centering
	\caption{Calculated and simulated Cherenkov photon yield per photoelectric interaction of a 511\,keV photon.}
	\label{tab:nbcherphotons}
		\begin{tabular}{lccc}
			\hline \hline
			&calculation&\multicolumn{2}{c}{simulation}\\
			Material&created photons&created photons&detected photons\\ \hline
			LSO:Ce&18&13.8&1.1\\
			LuAG:Ce&27&24.3&7.2\\
			BGO&28&32.8&4.6\\
			PWO&23&22.6\footnotemark[3]&3.8\footnotemark[3]\\ 
			Pb-glass&29&20.9\footnotemark[3]&3.3\footnotemark[3]\\ \hline \hline
		\end{tabular}
\end{table}

Comparing the numbers of created and detected Cherenkov photons in table \ref{tab:nbcherphotons} and figure \ref{fig:nbcherphotons}, reveals a high loss of Cherenkov photons during their propagation through the crystal to the photon detector. This loss is caused by photons leaving the crystal and, being the major factor, by photon absorption inside the crystal. The main reason for photon absorption is the high number of Cherenkov photons created with short wavelengths (proportional to $1/\lambda^2$ \cite{Jelley55}), where crystals are often not transparent, dependent on the lower cutoff frequency, $\lambda_1$. In the case of, e.\,g., LSO:Ce and also LuAG:Ce, many Cherenkov photons are absorbed in the range of the excitation bands due to the cerium doping. Excitation and emission bands overlap and, therefore, optical photons at these wavelengths are absorbed (self absorption) \cite{Ren04,Naud95}. This overlap and the influence of the cerium doping on the transmission spectrum for LuAG:Ce are illustrated in 
ref. \cite{Auffray09}. The absorption of Cherenkov photons could be decreased by adjusting (lowering) the amount of cerium doping. Increasing the Cherenkov yield with this method would lead to a decrease of the scintillation yield at the same time. Nevertheless, the total time resolution of the material can be improved, which will be shown in the following section.
 
\footnotetext[3]{Compton scattering is included.}

Comparing the numbers of the calculated Cherenkov photons in table \ref{tab:nbcherphotons} with ref. \cite{Lecoq10}, one notices a slight difference, which can be explained by the different wavelengths used for the calculations. Furthermore, the numbers for the detected photons in the same table show a large deviation from ref. \cite{Dolenec10}. This, however, can be explained mainly by the quantum efficiency used for the simulations and additionally by the dimensions of the crystals which is much larger in \cite{Dolenec10}, than in our case. Nevertheless, the interaction efficiency of the 511\,keV annihilation photons is proportional to the crystal lengths, which shows one trade-off in PET: a high interaction efficiency of the annihilation photon with the scintillator versus a high detection efficiency of optical photons.

\subsection{Influence of Cherenkov Photons on the Rise Time}
\begin{figure}[tb]
 \centering
\includegraphics[width=1\columnwidth,keepaspectratio]{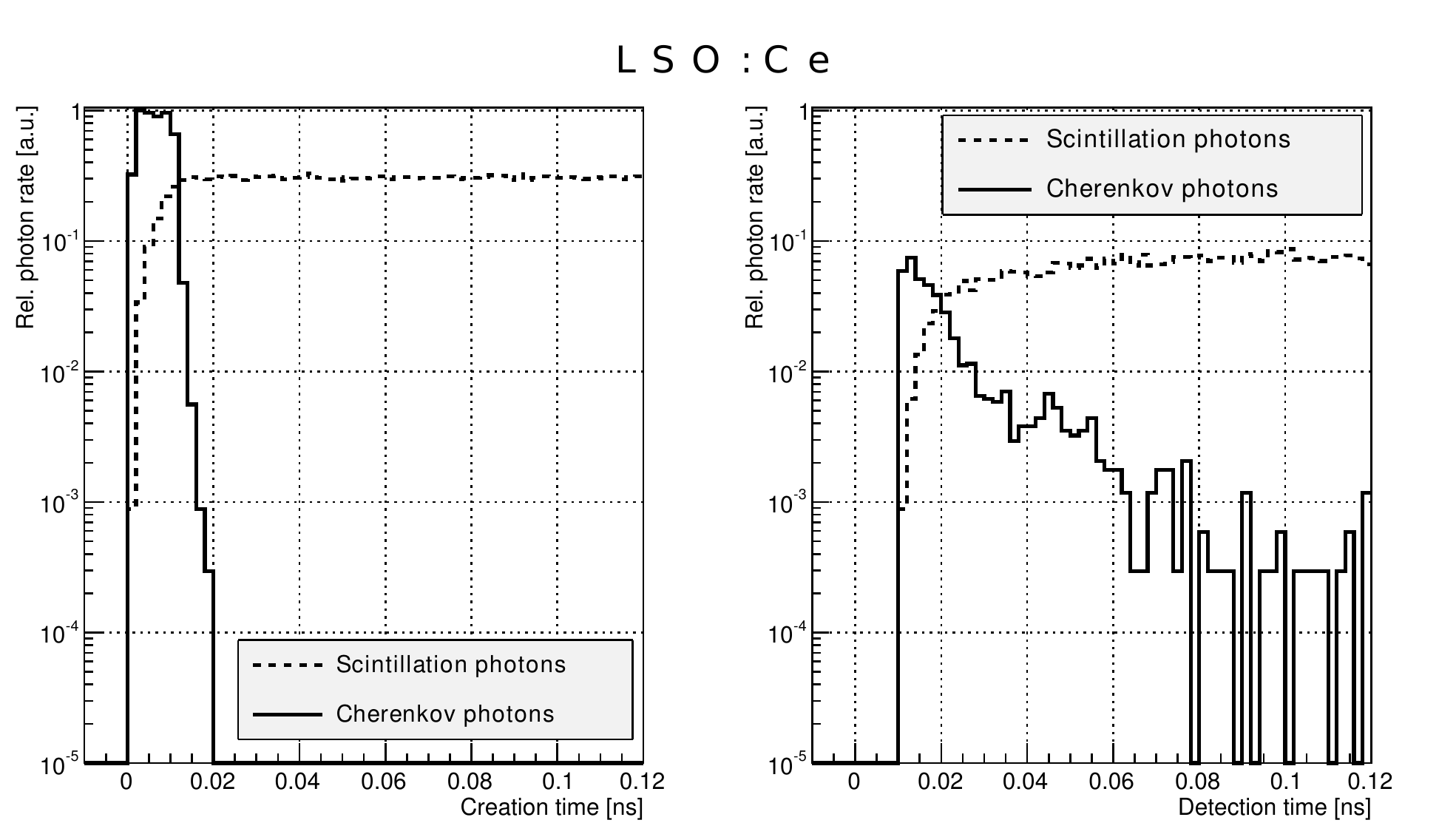}\\
\includegraphics[width=1\columnwidth,keepaspectratio]{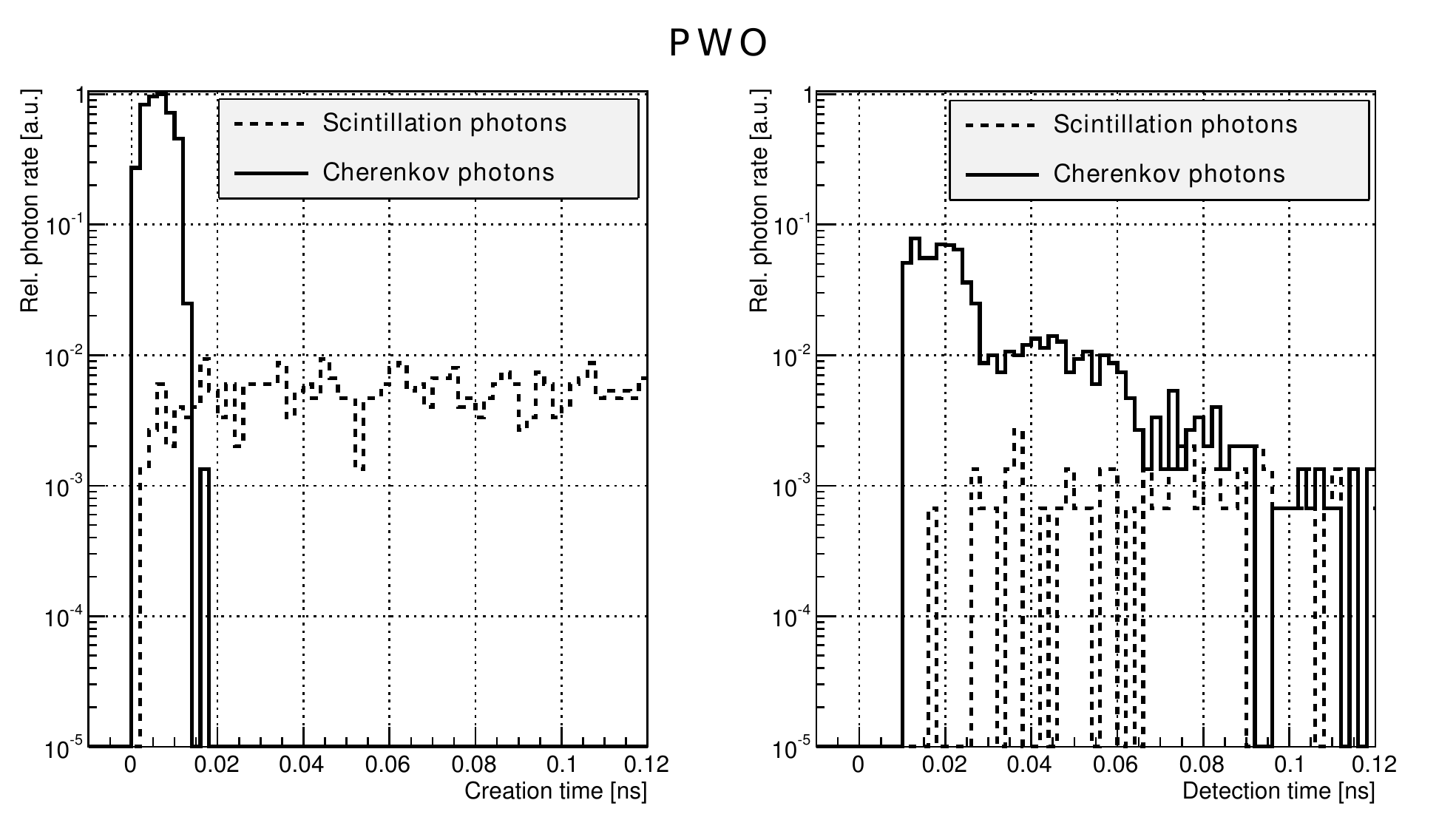}\\
\caption{Accumulated photon creation (left) and detection (right) rates at the photon detector for Cherenkov and scintillation photons for LSO:Ce (top) and PWO (bottom). The numbers of created and detected photons were normalized to the maximum creation rate of Cherenkov photons. A bin width of 2\,ps was chosen.}
\label{fig:risetime}
\end{figure}

A short rise time of scintillators is important for a good time resolution in TOF-PET \cite{Lecoq10, Moses99, Derenzo03} and can influence the CTR significantly \cite{Shao07}. As discussed above, the total yield of photons created in scintillators is composed of both, Cherenkov and scintillation photons. To investigate the influence of Cherenkov photons on the total rise time of scintillators, the photon creation rates for both, Cherenkov and scintillation photons have been simulated. The resulting rates are plotted in figure \ref{fig:risetime} for LSO:Ce and PWO, on the left side. The time distribution of photons arriving at the photon detector after propagating through the crystal can be seen on the right-hand side of figure \ref{fig:risetime}.

Although the size of the simulated scintillators is small, photon propagation inside the scintillators introduces a significant spread to the arrival times of the photons at the photon detector. These spreads can be seen when comparing the left and the right hand side of figure \ref{fig:risetime}. The accumulated photon creation and detection rates can be described by probability density distributions and contain information about the rise times of photon creation and the observable rise times at the photon detectors. 

Although the number of Cherenkov photons produced is low, their creation and detection rate can exceed the ones of scintillation photons, since they are created in a very narrow time span. This effect is visible in table \ref{tab:cherratio}, where a quantitative overview of the ratio of the number of Cherenkov photons and the number of scintillation photons within time windows of 25\,ps and 100\,ps is given. The time windows were measured from the time of creation or detection of the first photon, respectively. Considering a time window of 25\,ps only, the Cherenkov yields for all materials are exceeding those of the scintillation yields. These numbers suggest, that Cherenkov photons are an important factor for the rise times of scintillators. Depending on other scintillation parameters, e.\,g. the scintillation yields and transmission spectra, the influence of the Cherenkov photons on the rise times becomes more or less significant.

\begin{table}[tb]
\centering
	\caption{Simulated ratio of Cherenkov to scintillation yield.}%
	\label{tab:cherratio}
	\begin{tabular}{lcccc}
	  \hline \hline
	  &\multicolumn{4}{c}{Yield ratio}\\
	  &\multicolumn{2}{c}{created}&\multicolumn{2}{c}{detected}\\
	  Material&$<$\,25\,ps&$<$\,100\,ps&$<$\,25\,ps&$<$\,100\,ps\\ \hline
	  LSO:Ce&1.77&0.34&1.78&0.16\\
	  LuAG:Ce&11.5&2.1&41.5&3.4\\
	  BGO&122&24.2&364&28\\
	  PWO&86&16.6&134&21\\
	  \hline \hline 
	 \end{tabular}
\end{table}

As mentioned in the previous section, the ratio of Cherenkov and scintillation yield might be optimized by adjusting the amount of doping in some scintillators. Therefore, the time resolution of scintillators might be improved, simply due to a higher photon density at the beginning of the light pulse, even if the total light yield would be decreased.

\subsection{Angular Distribution}
\begin{figure}[tb]
 \centering
\includegraphics[width=1\columnwidth,keepaspectratio]{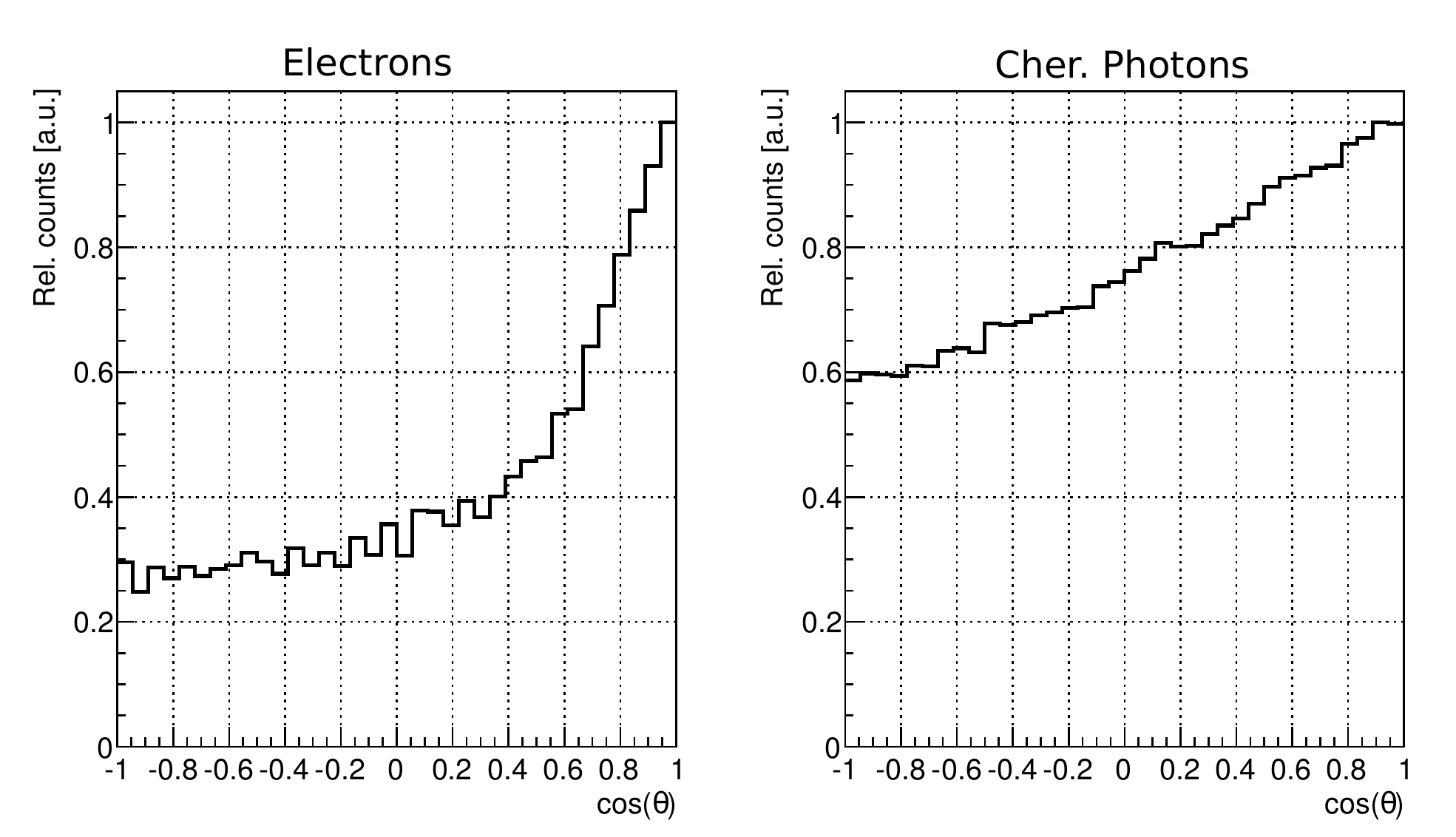}
\caption{Angular distribution of recoil electrons due to 511\,keV annihilation photons (left) and the subsequently emitted Cherenkov photons (right).}
\label{fig:angular_distribution}
\end{figure}

From table \ref{tab:nbcherphotons} it is visible, that the yield of Cherenkov photons at PET energies is low. In order to detect as many Cherenkov photons as possible, an optimized position for attaching photon detectors to Cherenkov radiators is important. Therefore, simulations on the angular distribution of Cherenkov photons have been performed for an LSO:Ce cube with 3\,mm edge length. For the axis of the spherical coordinate system, the flight direction of the incident 511\,keV photon was chosen to be the direction $\cos \theta = 1$, which, in the following, is also called forward direction.

In $\theta$-direction, for both, the electrons and the Cherenkov photons, an anisotropic distribution with a maximum at $\cos \theta = 1$ was observed, see figure \ref{fig:angular_distribution}. This behaviour and is stronger for the electrons than for the Cherenkov photons and could be used for optimizing the scintillator geometry and the positioning of the photon detectors on the crystals in order to maximize the detection yield of Cherenkov photons.

\section{Measurement}
Coincidence measurements have been performed to proof the principle of detecting Cherenkov photons due to the recoil electrons generated by the 511\,keV annihilation photons.

\begin{figure}[hbt]
 \centering
\includegraphics[width=0.85\columnwidth,keepaspectratio]{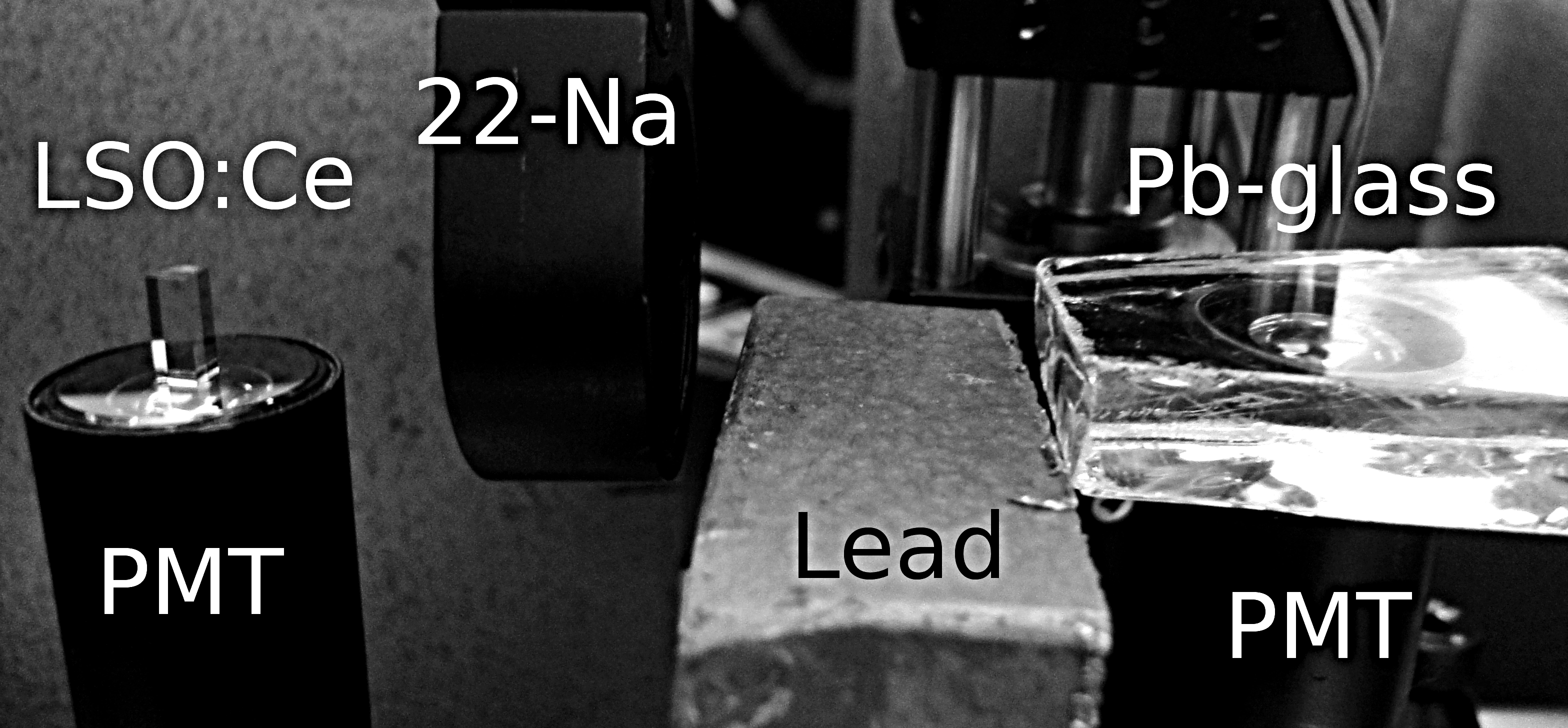}
\caption{Setup for the proof of principle of Cherenkov photon detection.}
\label{fig:leadglass}
\end{figure}

Two Hamamatsu R1450 PMTs with a transit time spread of 360\,ps (sigma) were used for the measurement. The PMTs were arranged in a coincidence setup with a $^{22}$Na source in the center, see figure \ref{fig:leadglass}. For optimizing the alignment, the source could be moved in vertical direction by a stepping motor. Artifacts due to 511\,keV photons entering the PMT and the PMT-window were avoided by placing a brick of lead in between the source and the PMT attached to the Cherenkov radiator. As Cherenkov radiator, lead glass RD50 from Schott, with a high fraction of lead-oxide ($>$ 65\%), with a size of $\sim$\,4\,cm\,$\times$\,5\,cm\,$\times$\,0.8\,cm and two faces polished was used. On the opposite side LSO:Ce, with a size of 3\,mm\,$\times$\,3\,mm\,$\times$\,8\,mm was coupled to the second PMT as reference detector. The output signals of both PMTs were split, with one part connected directly to a 4 channel digital WavePro 735Zi oscilloscope from LeCroy and the other part to a constant fraction 
discriminator (CFD, model 103, developed at PSI) before connected to the oscilloscope.

\begin{figure}[bt]
 \centering
\includegraphics[width=0.49\columnwidth,keepaspectratio]{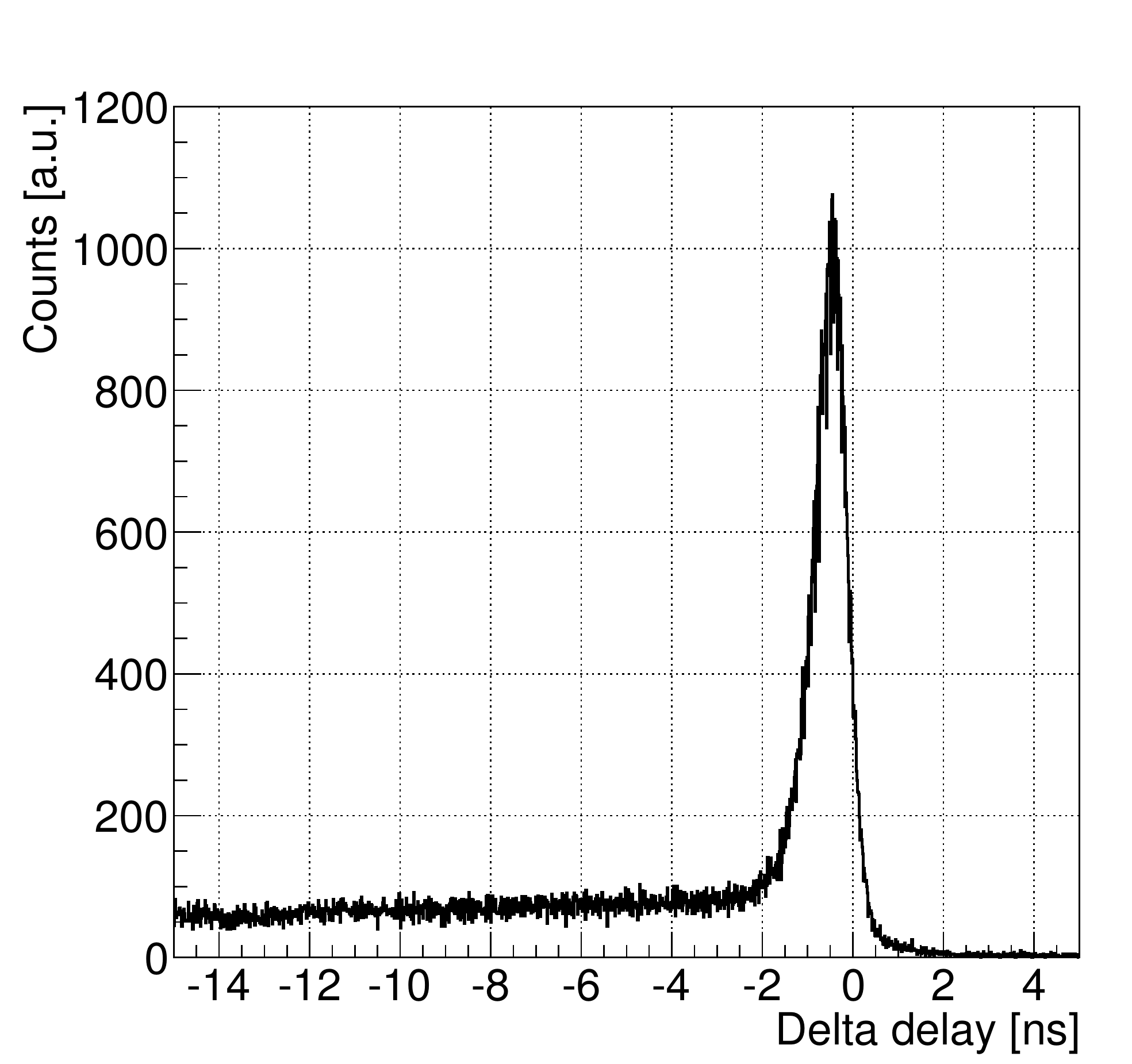}
\includegraphics[width=0.49\columnwidth,keepaspectratio]{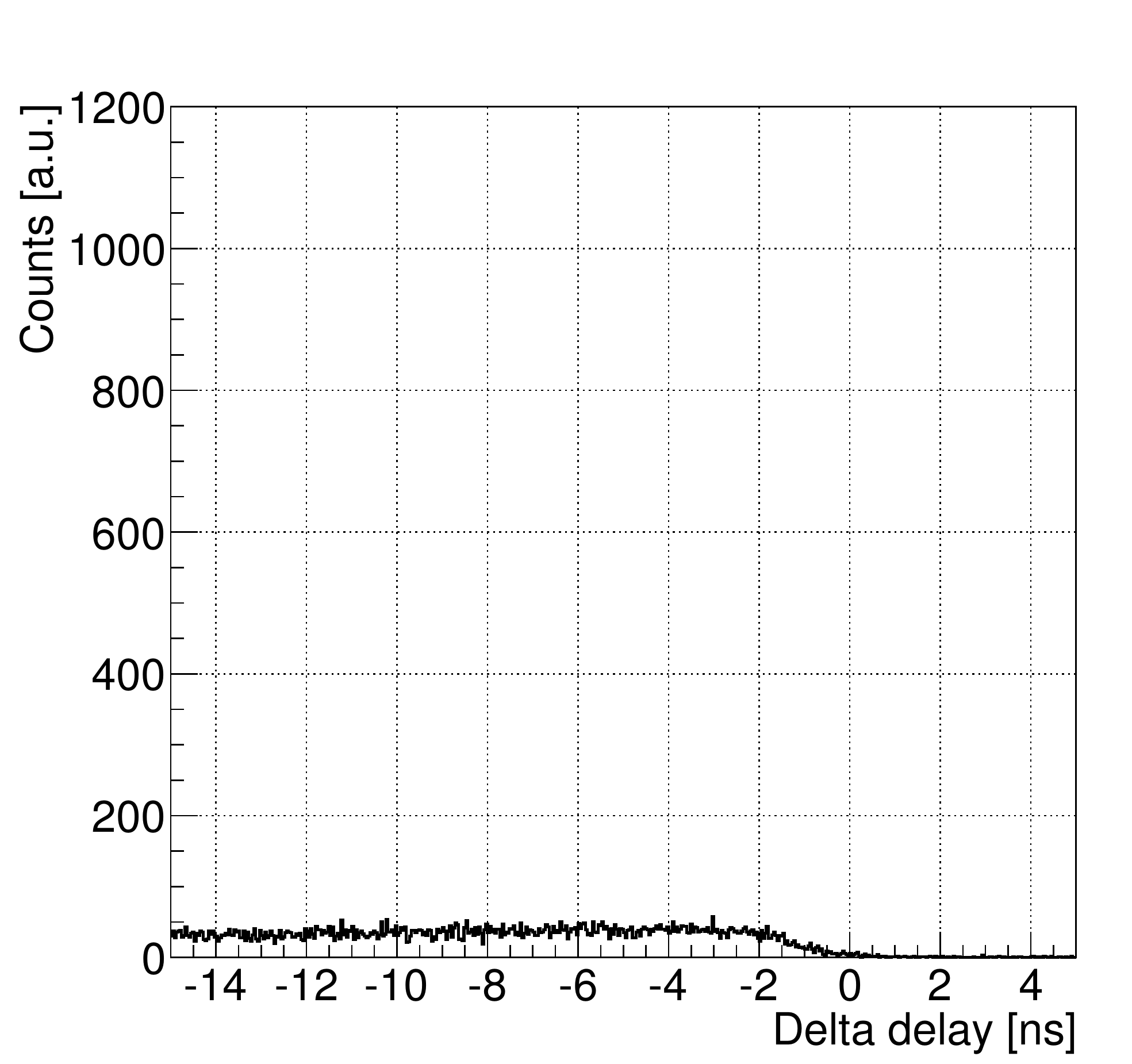}
\caption{Time difference of the two PMT signals from the coincidence measurement of lead glass vs LSO:Ce on the left. On the right, the measured background spectrum is shown, when the lead glass is removed from the PMT.}
\label{fig:leadglassvslso}
\end{figure}

For an accurate threshold setting, the output of the PMT with the lead glass on top was amplified using a NIM amplifier module 778 from Philips. The CFD thresholds were set to a level of 0.5 photons for the PMT attached to the lead glass and for the reference detector to a level between the Compton edge and the 511\,keV photo peak. The coincidence was done by triggering on the two CFD outputs and coincidence time resolution of 832\,ps FWHM was obtained, see figure \ref{fig:leadglassvslso} on the left side.

To ensure not to trigger on photons created in the PMT window, the measurement was repeated after removing the lead glass from the PMT. The obtained background spectrum is plotted in figure \ref{fig:leadglassvslso}, on the right-hand side. Comparison of the two plots proves that Cherenkov photons have been detected with this setup.

The relatively poor CTR is due to the equipment, which was chosen for a proof of principle only and not for achieving the best time resolution. By optimizing the setup, including the geometry of the lead glass, an improved CTR can be expected.

\section{Discussion}
For the investigated scintillators the Cherenkov yield is low compared to the scintillation yield. Nevertheless, during the first few ten picoseconds the emission rates of Cherenkov photons exceed the rates of scintillation photons for all investigated materials. This is even more obvious for scintillators like BGO and PWO due to their lower scintillation and higher Cherenkov yield, respectively. As the time windows, chosen for calculation of the Cherenkov yield, are in the range of the scintillation rise times, the Cherenkov effect seems to be an important factor influencing the total rise time of scintillators and thus, the time resolution of scintillation detectors could be improved.

Demonstrators using the Cherenkov effect only, already have been published and show promising results \cite{Dolenec10}. However, when using pure Cherenkov radiators, the energy resolution is poor and therefore, artifacts of PET images reconstructed from such data cannot be discriminated anymore.

Nevertheless, the Cherenkov effect can be exploited in combination with scintillation and, therefore, with sufficient energy resolution for PET. This might be achieved either by consecutive detection of Cherenkov and scintillation photons, or by the decrease of the rise time by increasing the Cherenkov-yield.

Consecutive readout would require very long rise times of the scintillation process and very fast photon detectors with low dark count rates in order to distinguish between Cherenkov and the subsequent scintillation photons. An advantage of this method would be the possibility to measure two time stamps (Cherenkov and scintillation) per event, which might improve the time resolution additionally.

Improvement of the time resolution by decreasing the rise time seems to be easier to realize. This can be done by optimizing the optical characteristics of scintillators in order to increase the Cherenkov yield and therefore, the photon density at the beginning of the scintillator emission response to a 511\,keV photon.

A major factor affecting the number of detected Cherenkov photons, is photon absorption inside the scintillator. Since the emission rate of Cherenkov photons is increasing at shorter wavelengths, materials with high transmission in the blue and UV-range, accompanied with photon detectors which are sensitive at these wavelengths are beneficial. Although photon detectors have a higher quantum efficiency at longer wavelengths, this cannot compensate the low yield of Cherenkov photons in this wavelength range and therefore would not lead to a significant increase of the Cherenkov detection yield.

Lead glass, usually used in radiation protection, was used for a proof-of-principle measurement, as it is a cheap and easy to get Cherenkov radiator which is free of scintillation. The equipment for the setup was chosen to detect Cherenkov photons with a basic setup, which explains the relatively poor coincidence time resolution of 832\,ps FWHM. Nevertheless, the detection of Cherenkov photons using lead glass is proven with this setup.

\section{Conclusion and Outlook}
The excellent timing properties of the Cherenkov photon emission could be exploited using materials with high Cherenkov yield, which might be achieved by optimizing the geometrical detector layout, refractive index and enhancing the transmission spectrum in the blue and UV-range. Eventually, this could lead to an improved total rise time of scintillators and therefore, to improved time resolution of TOF-PET. As a consequence, measurements of the rise times of inorganic scintillators, especially LSO:Ce or LuAG:Ce with varying Ce doping, would be very interesting, since a dependency of the rise time on the doping would strengthen the conclusions drawn from the simulations performed in this work.

\appendices







\bibliographystyle{IEEEtran}
%

%








\end{document}